\documentclass{pasa}%

\usepackage{graphicx}

\title[Resolving VLBI correlator ambiguity in the time delay model]{Resolving VLBI correlator ambiguity in the time delay model improves precision of geodetic measurements}

\author[O. Titov et al.]{O. Titov$^1$,  A. Melnikov$^{2}$, and Y. Lopez$^3$
\affil{$^1$Geoscience Australia, Canberra, Australia}%
\affil{$^2$Institute of Applied Astronomy of Russian Academy of Science, Saint-Petersburg, Russia}
\affil{$^3$University of Tasmania, Hobart, Australia}
}%

\jid{PASA}
\doi{10.1017/pas.\the\year.xxx}
\jyear{\the\year}

\usepackage{aas_macros}
\usepackage{hyperref} 
\hypersetup{colorlinks,citecolor=blue,linkcolor=blue,urlcolor=blue}

\begin{document}

\begin{frontmatter}
\maketitle

\begin{abstract}
The modern Very Long Baseline Interferometry (VLBI) relativistic delay model, as documented in the IERS Conventions refers to the time epoch when the signal passes one of two stations of an interferometer baseline (selected arbitrarily from the pair of stations and called the “reference station”, or "station 1"). This model consists of the previous correlation procedure used before the year 2002. However, since 2002 a new correlation procedure that produces the VLBI group delays referring to the time epoch of signal passage at the geocenter has been used. A corresponding correction to the conventional VLBI model delay has to be introduced. However, this correction has not been thoroughly presented in peer reviewed journals, and different approaches are used at the correlators to calculate the final group delays officially published in the IVS database. This may cause an inconsistency up to 6 ps for ground-based VLBI experiments between the group delay obtained by the correlator and the geometrical model delay from the IERS Conventions used in data analysis software. Moreover, a miscalculation of the signal arrival moment to the "reference station" could result a larger modelling error (up to 50 ps). The paper presents the justification of the correction due to transition between two epochs elaborated from the Lorentz transformation, and the approach to model the uncertainty of the calculation of the signal arrival moment. The both changes are particularly essential for upcoming broadband technology geodetic VLBI observations.

\end{abstract}

\begin{keywords}
IVS -- broadband Very Long Baseline Interferometry (VLBI) -- relativity -- Geodesy -- Lorentz transformation -- reference radio sources
\end{keywords}
\end{frontmatter}

\section{INTRODUCTION }
\label{sec:intro}

The Very Long Baseline Interferometry (VLBI) technique measures the difference between the arrival times of a signal from a distant radio source at two radio telescopes (\cite{Schuh12}). The signal is recorded at each radio telescope together with time marks from independent hydrogen masers. Due to separation of the radio telescopes by a few hundred or thousand kilometres, the plain  wave front passes first telescope earlier then the second one. This difference in the arrival time of the signal at both radio telescopes is known as time delay, and the frequency shift due to the relative motion of the telescopes around the geocentre is known as delay rate. 

The time delay and delay rate are found during cross-correlation of the two independent records. There are two types of correlators (XF and FX) based on the order of the mathematical operations – cross-correlation (X) and Fourier transformation (F). Baseline-based correlators are designed as XF type correlators, and station-based correlators are FX type correlators. For the baseline-based XF-type MarkIII correlator used before 2002, the observables referred to the position of one of the two stations (station 1). For the station-based FX-type MarkIV correlator all observables for all baselines at one single multi-baseline scan are referred to the geocentre as a common reference point. As 1 ps precision is required for the time delay calculation, all first-order and second-order effects of special relativity should be taken into account.

One of the goals of the International VLBI Service activities is to achieve 1-mm accuracy from the analysis of routine geodetic VLBI sessions. The accuracy of the daily scale factor improved dramatically in 2002 when the MarkIII correlator was replaced by MarkIV correlator (\cite{2018A&A...610A..36T}). However, so far this value varies about 3-4 mm despite technological developments since 2002. 

One possible reason for the lack of improvement in accuracy is the inconsistency between the VLBI observable group delays and the relativistic delay model developed in 1980s-90s, published in the IERS Conventions 2010 (\cite{iers10}). The transition from the MarkIII to MarkIV correlator was not followed by any changes in the IERS Conventions model that still refers to the epoch of the wavefront passage of station 1. Thus, it remains consistent with the XF-type correlators. To make the output delay of the FX-correlator consistent with the IERS Conventions 2010 model (XF-type), an additional correction needs to be applied. Unfortunately, this correction has not been officially presented in explicit form. This conversion difference was called “subtle” (\cite {2000ivsg.conf..187W}); however, it reaches 20 ns, which is quite significant. \cite{Corey2000} developed a simple geometric approach under assumption of the finiteness of the speed of light to obtain this correction, but his final equation comprised a major term only, while several minor terms were not included.

In this paper, it is emphasised that the relativistic correction due to the change of the reference epoch definition should be derived from the Lorenz transformation to secure the 1-ps accuracy. Therefore, the final equation of the recommended group delay should include some minor relativistic terms due to coupling of the barycentric and geocentric velocities
of the radiotelescopes to be added to the version by \cite{Corey2000}. A detailed development of the correction based on the Lorenz transformation is given in Appendix. This correction is essential from the theoretical point of view; however, its impact on the geodetic results is negligible for ground-based baselines (less than 1 mm). 

A more serious problem is caused by the uncertainty in the signal arrival time as calculated by the correlators, even if the problem of the epoch calculation is fixed. Within the adopted procedure, for a single multi-station scan this time is common for all stations and is usually rounded to integer number of seconds. Meanwhile, for a multi-station scan, the factual signal arrival time is individual for each station, the output group delay is converted to the time common for all stations within one scan using a reasonable polynomial approximation. Therefore, the final output delay for each baseline is referred to the common time of scan. Theoretically, this output delay should be perfectly consistent to the delay at the time of the signal arrival to the "reference station" of each baseline. However, this is not guaranteed due to the uncertainty of the reference epoch definition (discussed in the Appendix) and hidden numerical issues during the polynomial approximation. 

To estimate an additional correction, the standard parametrical model should be extended. For each scan we have a time of the signal arrival (common for $N$ stations) and a set of $N(N-1)/2$ time delays for all baselines. Instead of seeking for $N(N-1)/2$ errors in the delays themselves, it would be easier to treat the signal arrival time as the parameter to be updated assuming the delays are errorless. A possible approach to model this type inconsistency is presented analytically in \cite{Finkelstein1983}. A second order term in Equation (\ref{delta_tau}) may be generalised at 1-ps accuracy in the form 

\begin{equation}\label{epsilon_tau}
\begin{aligned}
\delta{\tau_{12}} =  \frac{(\boldsymbol{b}\cdot\boldsymbol{s})}{c^2}
\frac{(({\epsilon\boldsymbol{w_1}+\boldsymbol{w_2})\cdot\boldsymbol{s})}}{1+\epsilon}\\
\end{aligned}
\end{equation}

The case $\epsilon = 0$ corresponds to the selection of the reference clock at station 1, and the case $\epsilon = \infty$ corresponds to the selection of the reference clock at station 2. The relativistic group delay model from the IERS Conventions has an intrinsic assumption that $\epsilon = 0$. A violation of this assumption results in a small deviation of the $\epsilon$ from zero. For a small value of $\epsilon$ it could be parametrized with the partial derivative

\begin{equation}\label{partial_epsilon_tau}
\begin{aligned}
\frac{\partial\delta{\tau_{12}}}{\partial\epsilon} =  \frac{(\boldsymbol{b}\cdot\boldsymbol{s})
(({\boldsymbol{w_1}-\boldsymbol{w_2})\cdot\boldsymbol{s})}}{c^2}\\
\end{aligned}
\end{equation}

By its analytical representation, this new parameter $\epsilon$ should be referred to the group of parameters to model the clock instability (offset, rate, 2nd derivative, etc). In total, $(N-1)$ parameters should be added to the traditional procedure of the VLBI delay modelling.

The parameters $\epsilon$ could be estimated with Equation (\ref{partial_epsilon_tau}) by the least squares individually for each VLBI station clock except to the clock at the network "reference station" that is assumed to be errorless.  Then for two arbitrary stations (i and j) the corresponding delay is calculated as follows

\begin{equation}\label{epsilon_tau_ij}
\begin{aligned}
\delta{\tau_{ij}} =  (\epsilon_{i} - \epsilon_{j})\frac{(\boldsymbol{b_{ij}}\cdot\boldsymbol{s})
((\boldsymbol{w_i}-\boldsymbol{w_j})\cdot\boldsymbol{s})}{c^2}\\
\end{aligned}
\end{equation}

\section{DATA ANALYSIS}

\begin{figure}
\begin{center}
\includegraphics[width=20pc, height=15pc]{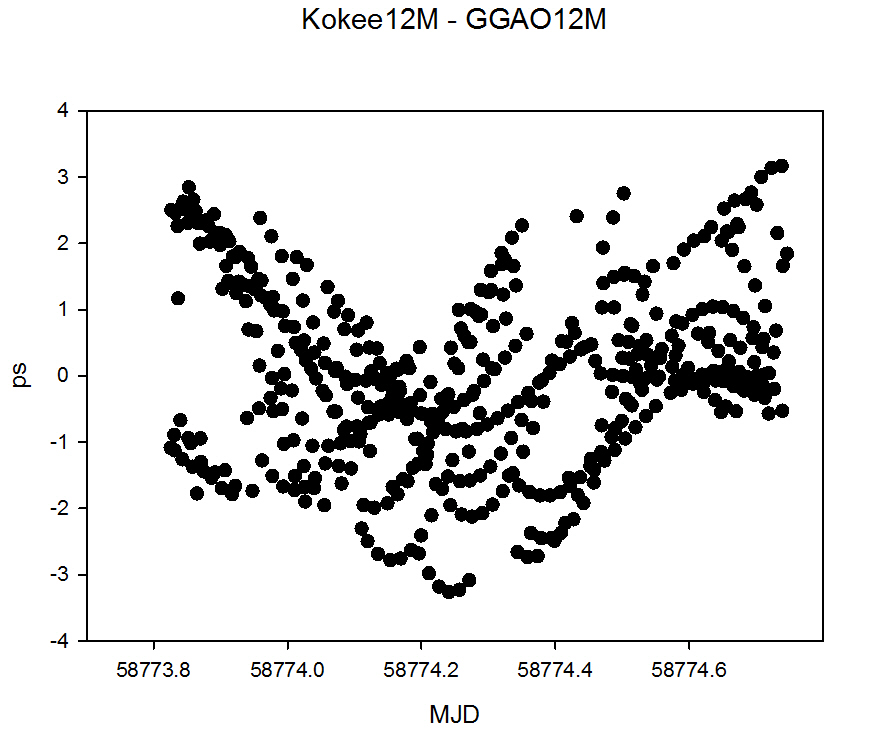} 
\vspace{0.1 cm}
\includegraphics[width=20pc, height=15pc]{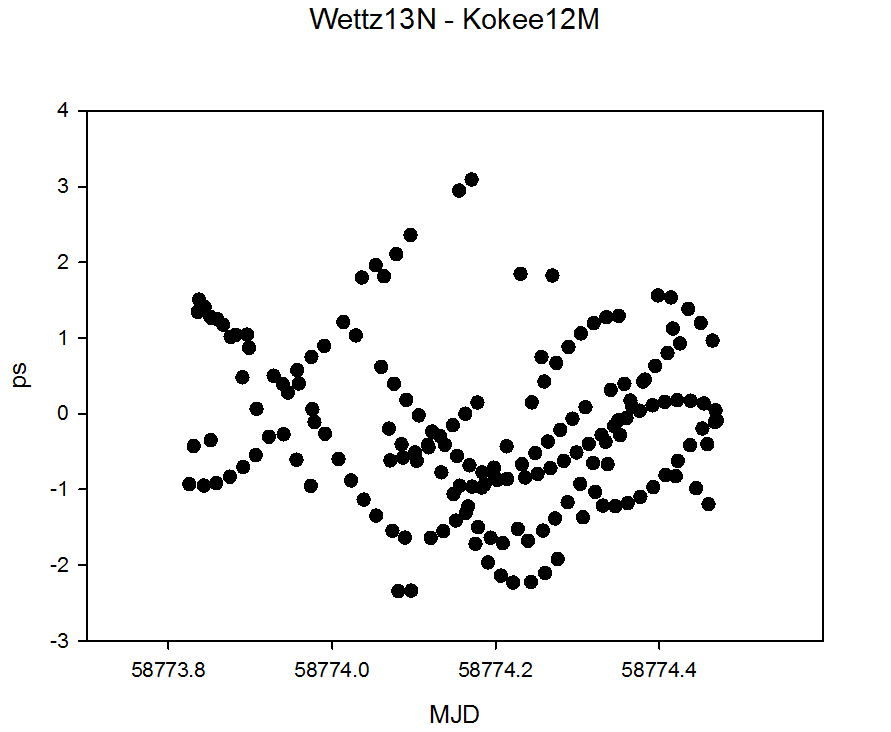}
\caption{Contribution of the three third order terms from Eq (11) for baselines KOKEE12M - GGAO12M (7405 km) (top) and WETTZ13S - KOKEE12M (10358 km) (bottom).}
\label{Fig1}
\end{center}
\end{figure}

 The second term in Equation (\ref{delta_tau}) (of the Appendix) is the diurnal variation of the Earth scale's factor that replaces the diurnal aberration applied for the traditional astronomical observations. This is the only term due to the Earth's rotation implemented by the FX-correlator software developers (in accordance to \cite{Corey2000}). However, a more accurate approach based on the Lorenz transformation (\ref{xt3}) reveals additional minor terms in Equation (\ref{delta_tau}) due to coupling of the two velocities $V$ and $w_2$. The first term in Equation (\ref{delta_tau}) is the coordinate term due to the transformation from the barycentric to the geocentric reference frame, and it could be ignored for the scope of this paper.

\begin{figure}
\begin{center}
\includegraphics[width=20pc, height=15pc]{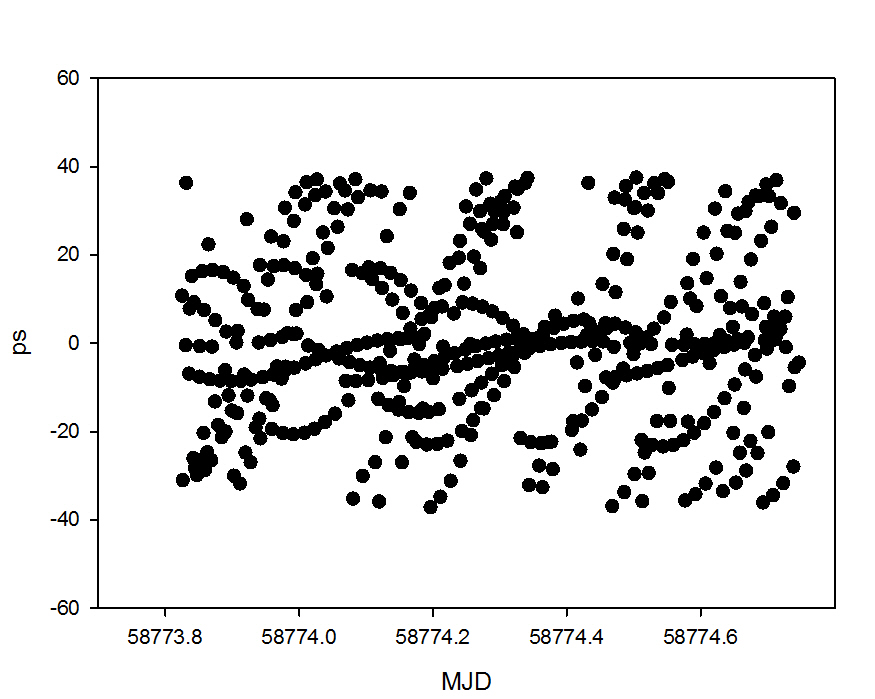}
\vspace{0.1 cm}
\caption{Systematic group delay for baseline KOKEE12M - GGAO12M (7405 km) in accordance with 
(\ref{epsilon_tau_ij}).}
\label{Fig2}
\end{center}
\end{figure}

We used one of the recent VGOS experiments (VT9290, 17-Oct-2019) for more detailed analysis. This 24-hour experiment included five radio telescopes (WETT13S, ONSA13NE, ONSA13SW, GGAO12M and KOKEE12M) equipped with the broad band receivers. Observations were performed in four bands with dual linear polarisation (3000-3480 MHz, 5240-5740 MHz, 6360-6840 MHz and 10200-10680 MHz) (\cite{Alef2019}). 
 

Fig \ref{Fig1} shows the contribution of the three “missed” terms in Equation (\ref{delta_tau}) to the total delay for two baselines: KOKEE12M - GGAO12M (7405.4 km) and KOKEE12M - WETTZ13S (10357.6 km). As expected, the correction on Fig \ref{Fig1} is essential for long baselines (up to 6 ps).

Standard geodetic VLBI observations operated in two frequencies, 2.3 GHz (S-band) and 8.4 GHz (X-band), are not sensitive to the effect of the time of signal arrival. Therefore, we used the new broad band VLBI data (VGOS project) to estimate the parameter $\epsilon$. Due to the higher sample rate and the broader bandwidth of the recorded data the formal accuracy of the VGOS geodetic results is better than for standard S/X observations by an order of magnitude.

The VGOS data files were processed using the OCCAM software (\cite{Titov2004}) (version 6.3) in two modes. A first solution produces a standard set of parameters for estimating - (i) corrections to the positions of radio telescopes in the ITRF2014 frame (\cite{Altamimi_2016}), (ii) Earth orientation parameters, (iii) wet troposphere delay and two gradients, (iv) three parameters to model the clock instability for each station except the reference one (clock offset, clock rate and second derivative), and (v) corrections to the ICRF3 positions of several radio sources that expose a high level of astrometric instability in the past. 
A second solution was used to estimate the parameter $\epsilon$ for all stations except for the reference one. 

Estimates of the parameter $\epsilon$ for six VGOS stations operating during 2019 are shown in Table 1. About half of the estimates are statistically significant. This means that typically, the time of the radio wave arrival to the reference station is not calculated by the correlator with sufficient accuracy. The resulting group delay calculated by Equation (\ref{epsilon_tau_ij}) for baseline GGAO12M - KOKEE12M at the same session (17-Oct-2019, MJD = 58744) is shown in Fig \ref{Fig2}. We selected this baseline because for both stations in this experiment the estimates of $\epsilon$ are larger than usual ($-0.885\cdot10^{-3}$ for GGAO12M and $0.892\cdot10^{-3}$ for KOKEE12M). The range of the peak-to-peak variations is about 80 ps. This results in additional, hidden, source of systematic error for all other parameters.

\subsection{ANALYSIS OF ASTROMETRIC RESULTS}

A comprehensive analysis of geodetic parameters is beyond of the scope of this paper. Herewith we discuss only effect of the additional parameter on the astrometric positions of two well-known reference radio sources, namely 0552+398 and 1156+295. Both sources were observed in twenty 24-h broadband VLBI experiments during 2019, with a large number of observations. As a result, their formal positional errors for both components are less than 50 $\mu$as for almost all experiments. Therefore, statistical investigation of the astrometrical results would demonstrate the advantage of the new VLBI technology application and the effect of the inclusion of the additional modelling parameter.

\subsubsection{Radio source 0552+398}

Radio source 0552+398 is one of the most actively observed radio sources by geodetic VLBI since 1979 due to its strong flux density and good astrometric stability. It was included to the list of reference radio sources of ICRF1, (\cite{Ma1998}), ICRF2 (\cite{Fey_2015}) and ICRF3 (\cite{Charlot_2020}). It was also treated as a 'stable' one after independent verification by \cite{Feissel-Vernier_2003}. The source 0552+398 has no apparent structure at S- and X-bands images. However, its imaging at higher frequencies (24 and 43 GHz) discloses a sub-milliarcsec jet in the east direction from the core (\cite{Charlot2010}). Recently, a second component was revealed by \cite{Bolotin_2019} from the analysis of the broadband observations during the CONT17 campaign.

\begin{figure}
\begin{center}
\includegraphics[width=18pc, height=12pc]{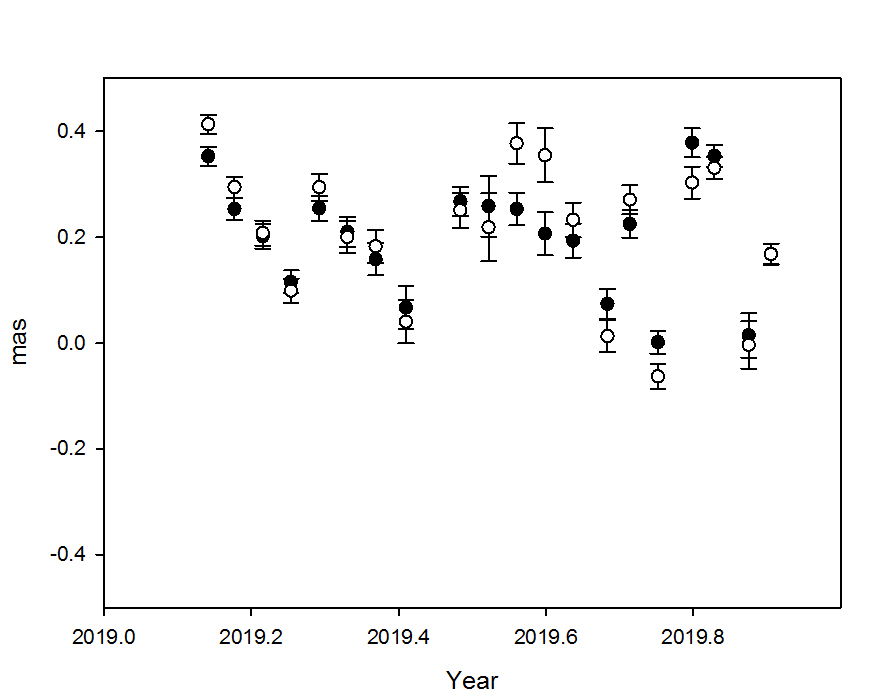} 
\vspace{0.1 cm}

\includegraphics[width=18pc, height=12pc]{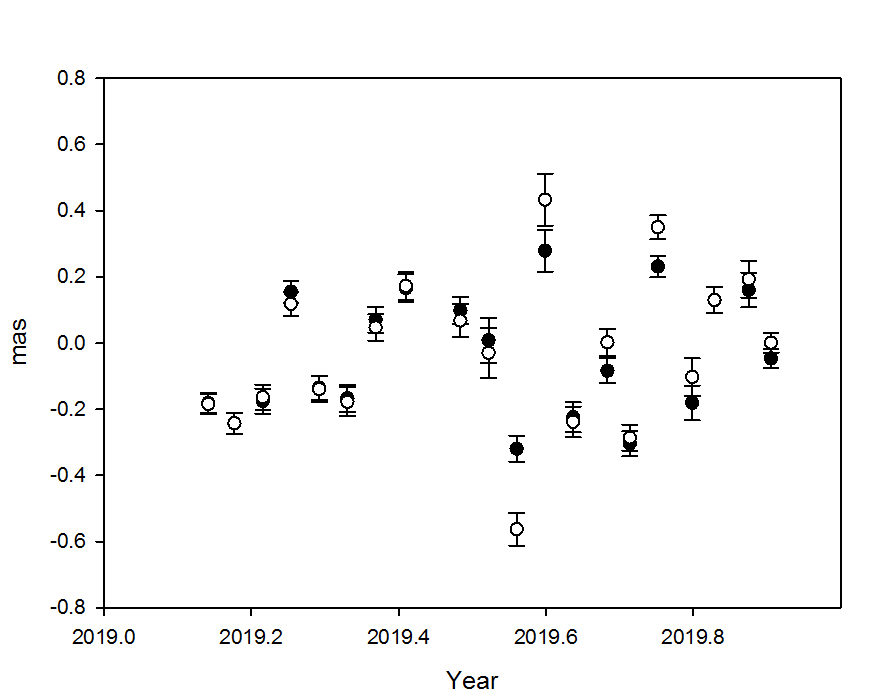}
\caption{Daily corrections to the ICRF3 coordinates of the radio source 0552+398 (up: right ascension bottom: declination). Black circles - standard solution, white circles - solution included the parameter $\epsilon$.}
\label{Fig3}
\end{center}
\end{figure}

While the daily estimates of the corrections to the declination component in Fig \ref{Fig3} vary around the original ICRF3 catalogue position within 0.6 mas, the estimates of the correction to right ascension (RA = 05h 55m 30s.80561207) (\cite{Charlot_2020}) show a non-zero offset of approximately 0.2 mas. The available post-ICRF3 catalogues (e.g. the celestial reference frame solution aus2020a published by International VLBI Service (IVS)) including S/X observations during 2019-2020 do not detect any essential offset of the 0552+398 positions with respect to the ICRF3 catalogue coordinates. This potentially indicates that the jet observed at high frequencies (24 and 43 GHz) is also essential for frequencies between 2 and 11 GHz, even though it is not detected on the S/X images. We conclude that the broadband VLBI observations are more sensitive to the sub-milliarcsec structure than the traditional S/X VLBI observations as also hinted by \cite{Bolotin_2019}. Therefore, the positions of the reference radio sources observed by the new broadband technology are not necessary to be coincided with the S/X data positions. 

\begin{figure}
\begin{center}
\includegraphics[width=18pc, height=12pc]{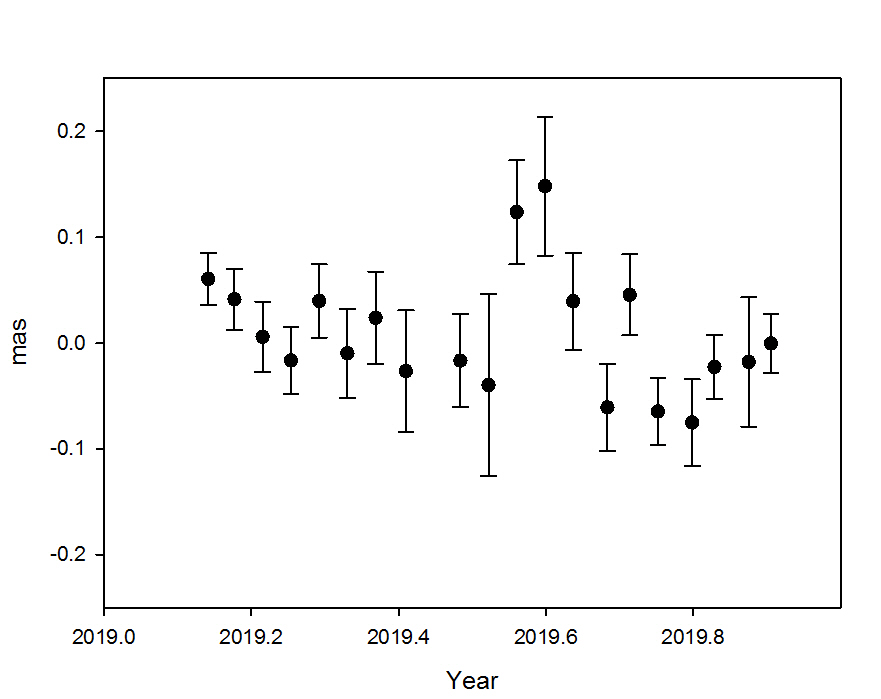} 
\vspace{0.1 cm}

\includegraphics[width=18pc, height=12pc]{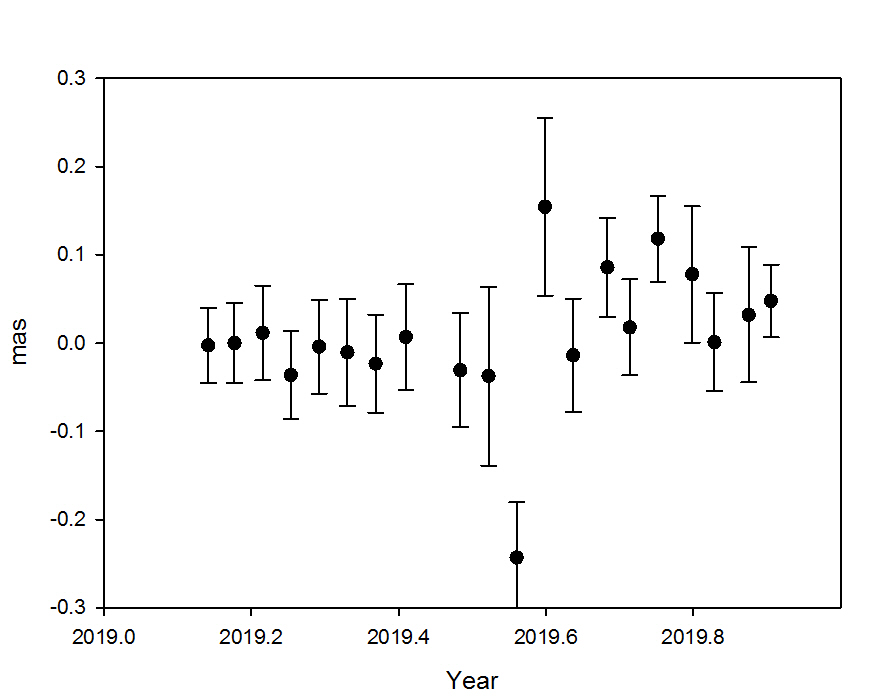}
\caption{Difference between two solution corrections for the radio source 0552+398 (up: right ascension bottom: declination).}
\label{Fig4}
\end{center}
\end{figure}

  

\subsubsection{Radio source 1156+295}

\begin{figure}
\begin{center}
\includegraphics[width=18pc, height=12pc]{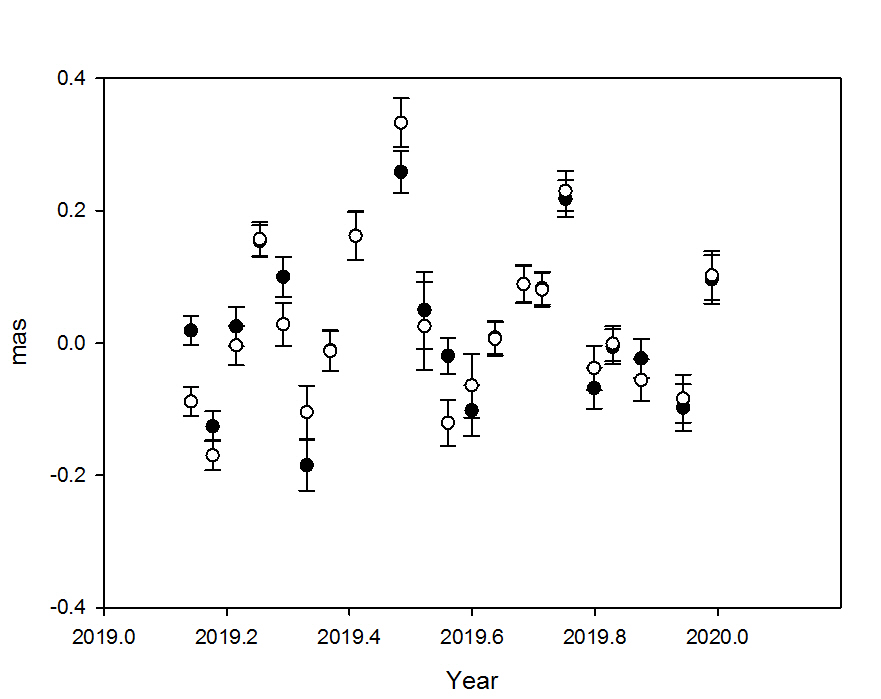} 
\vspace{0.1 cm}

\includegraphics[width=18pc, height=12pc]{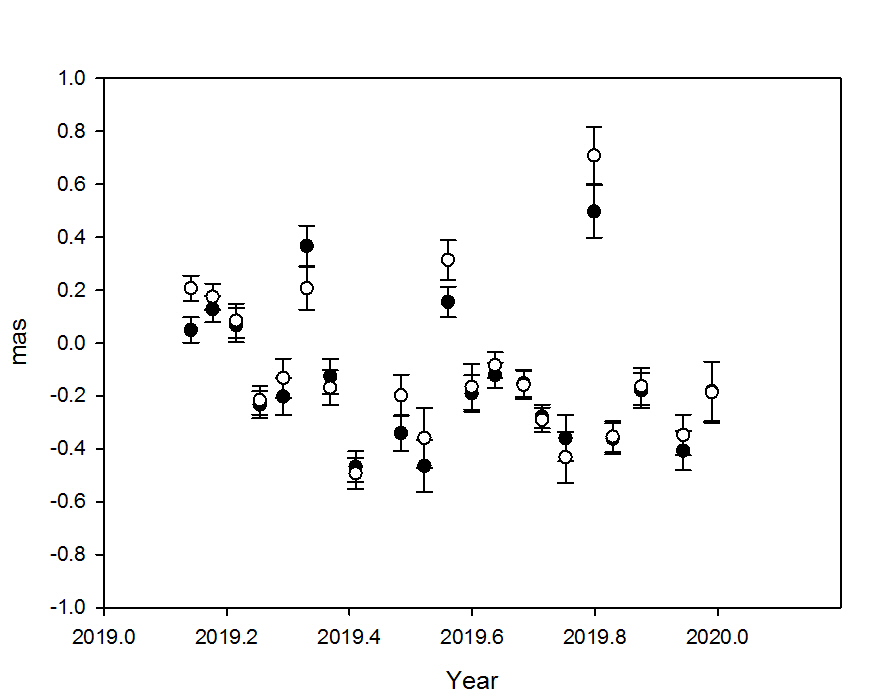}
\caption{Daily corrections to the ICRF3 coordinates of the radio source 1156+295 (up: right ascension bottom: declination). Black circles - standard solution, white circles - solution included the parameter $\epsilon$.}
\label{Fig5}
\end{center}
\end{figure}

Radio source 1156+295 has actively monitored for the last 30 years over a wide range of frequencies. Despite its extended structure in S- and X-bands with an elongated jet in the north direction (e.g. \cite{Kellermann1998}), the radio source 1156+295 demonstrates a moderate range of astrometric instability. At the same time, no structure was reported in 24 GHz and 43 GHz (\cite{Charlot2010}). Therefore, it was selected as one of the defining reference sources in the second ICRF realization (ICRF2) (\cite{Fey_2015}), although, not included to the list of the ICRF3 reference sources. 
Our analysis of the broadband VLBI results highlights a higher range of astrometric instability in declination than right ascension time series (Fig \ref{Fig5}) during 2019, presumably, induced by the jet oriented to the north direction. The average declination component is shifted approximately 0.2 mas south with respect to the ICRF3 catalogue position.

\begin{figure}
\begin{center}
\includegraphics[width=18pc, height=12pc]{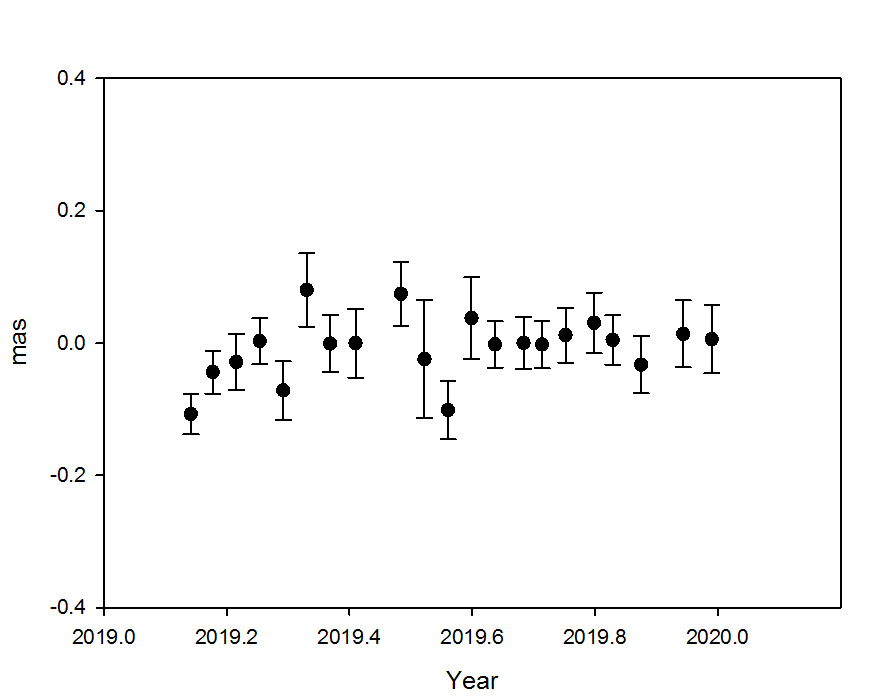} 
\vspace{0.1 cm}

\includegraphics[width=18pc, height=12pc]{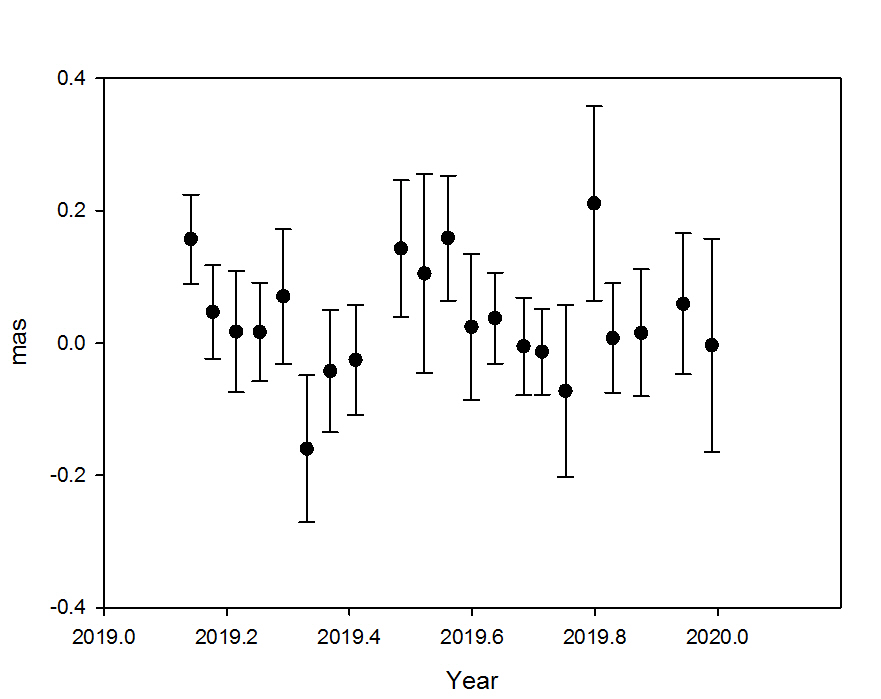}
\caption{Difference between two solution corrections for the radio source 1156+295 (up: right ascension bottom: declination).}
\label{Fig6}
\end{center}
\end{figure}

The difference between the two sets of daily estimates in Fig \ref{Fig4} and \ref{Fig6} does not reveal any noticeable astrometric signature due to inclusion of $\epsilon$ to the list of estimated parameters. For both sources the peak-to-peak variations do not exceed 0.25 mas in both components. Therefore, for radio sources 0552+398 and 1156+295, the inclusion of new parameter does not change the source position estimates essentially. However, for rare observed radio sources this difference may cause a substantial change in the final catalogue positions. 

\section{DISCUSSION AND CONCLUSION}

The transition from the XF-type to FX-type correlators for processing geodetic VLBI data requires a corresponding revision of the relativistic group delay in the IERS Conventions to secure a match between the correlator output and the theoretical model. Alternatively, a special correction needs to be done at the final step of the post-correlation data processing. In Equation (\ref{delta_tau}) we show in the four last terms the relativistic correction due to the time transformation from the epoch of the geocenter to the epoch of station 1. This correction is derived from the modified version of the Lorenz transformation in Equation (\ref{xt3}). Missing of the three minor terms in Equation (\ref{delta_tau}) can lead to a discrepancy of the group delay model at a level of 6 ps for long baselines. This is, in particular, pertinent for the intensive experiments for rapid estimation of Universal Time because a typical observational network consists of 2 or 3 radio telescopes separated by a long baseline (> 7000 km). We would like to recommend this equation be applied for the post-processing analysis of VLBI data at the modern FX-correlators.

Another effect, though may not be directly linked to the first one, is the uncertainty of the time of signal registration for each telescope as measured by the local clock (hydrogen maser) at the reference station and extrapolated during the process of correlation. This effect also refers to the difference of the geocentric velocities of the both radio telescopes, but it could be introduced as the extension of the clock instability model. The additional parameter describes how far the actual time of the signal arrival deviates from the time presented in the VLBI data file. Our analysis of broadband VLBI data over 2019 reveals that the parameter is statistically significant in many cases (Table 1). The corresponding systematic effect is up to 100 ps in time delays, and up to 0.25 mas in estimates of daily radio source positions. 

It is not yet clear whether the source structure effect directly links to the problem of precisely determining the time of the signal arrival to the radio telescopes. The algorithm of the numerical calculation of the signal arrival time always relies on the assumption that the phase reference point of the target source is the same for all frequency bands. However, with a broadband receiver, we may have four different phase reference points at the four frequency bands. Therefore, four signals in each band may arrive to the receiver at four different times even from a point-like radio source. A standard calibration may not compensate this inconsistency perfectly, mostly due to the non-linear behaviour of the phase during the fringe-fitting process. In addition, an extended radio source may have four different phase reference points at four frequencies referring to the celestial reference frame. Thus, the actual differences between the signal arrival times for four frequency bands could change unpredictably. As a result, the signal arrival time presented in the broadband VLBI data file as a single value has some level of uncertainty making the additional parameter $\epsilon$ feasible for routine application using Equation (\ref{epsilon_tau_ij}). While it was not essential for the traditional S/X VLBI observations, the broadband VLBI observations are more accurate, and, more advanced parametrical model should be used to match these observations.

\begin{acknowledgements}
We are thankful to Sergei Kopeikin (University of Missouri-Columbia), Slava Turyshev (JPL), James Anderson (GFZ), and Igor Surkis (IAA RAS) for fruitful discussions on the theoretical aspects and the technical details of the correlation process.
Also we thank the PASA Editor-in-Chief, and the anonymous referee for their constructive comments and suggestions which have significantly improved the clarity of the paper.

This paper is published with the permission of the CEO, Geoscience Australia.

We used the International VLBI Service (IVS) products available electronically at http://ivscc.bkg.bund.de/products-data/products.html.

\end{acknowledgements}

\bibliographystyle{pasa-mnras}
\bibliography{oleg}

\begin{thebibliography}{26}
\providecommand{\natexlab}[1]{#1}
\providecommand{\url}[1]{{#1}}
\providecommand{\urlprefix}{URL }
\expandafter\ifx\csname urlstyle\endcsname\relax
  \providecommand{\doi}[1]{DOI~\discretionary{}{}{}#1}\else
  \providecommand{\doi}{DOI~\discretionary{}{}{}\begingroup
  \urlstyle{rm}\Url}\fi
\providecommand{\eprint}[2][]{\url{#2}}

\bibitem[\protect\citeauthoryear{Alef et al.}{2019}]{Alef2019} Alef W., Anderson J.~M., Bernhart S., de Vicente P., Gonz{\'a}lez Garc{\'\i}a J., Haas R., La Porta L., et al., 2019, evga.conf, 24, 107

\bibitem[\protect\citeauthoryear{Altamimi et al.}{2016}]{Altamimi_2016} Altamimi Z., Rebischung P., M{\'e}tivier L., Collilieux X., 2016, JGRB, 121, 6109. doi:10.1002/2016JB013098

\bibitem[\protect\citeauthoryear{Bolotin et al.}{2019}]{Bolotin_2019} Bolotin S., Baver K., Bolotina O., Gipson J., Gordon D., Le Bail K., MacMillan D., 2019, evga.conf, 24, 224

\bibitem[\protect\citeauthoryear{Charlot et al.}{2010}]{Charlot2010} Charlot P., Boboltz D.~A., Fey A.~L., Fomalont E.~B., Geldzahler B.~J., Gordon D., Jacobs C.~S., et al., 2010, AJ, 139, 1713. doi:10.1088/0004-6256/139/5/1713

\bibitem[\protect\citeauthoryear{Charlot et al.}{2020}]{Charlot_2020} Charlot P., Jacobs C.~S., Gordon D., Lambert S., de Witt A., B{\"o}hm J., Fey A.~L., et al., 2020, Astron.Astroph. (in press), arXiv, arXiv:2010.13625

\bibitem[\protect\citeauthoryear{Corey}{2000}]{Corey2000} Corey B., 2000, Memo of Massachusets Institute of Technology, Haystack Observatory

\bibitem[\protect\citeauthoryear{Feissel-Vernier}{2003}]{Feissel-Vernier_2003} Feissel-Vernier M., 2003, Astron.Astroph., 403, 105. doi:10.1051/0004-6361:20030348

\bibitem[\protect\citeauthoryear{Fey et al.}{2015}]{Fey_2015} Fey A.~L., Gordon D., Jacobs C.~S., Ma C., Gaume R.~A., Arias E.~F., Bianco G., et al., 2015, AJ, 150, 58. doi:10.1088/0004-6256/150/2/58

\bibitem[\protect\citeauthoryear{Finkelstein, Kreinovich, \& Pandey}{1983}]{Finkelstein1983} Finkelstein A.~M., Kreinovich V.~I., Pandey S.~N., 1983, ApSS, 94, 233. doi:10.1007/BF00653714

\bibitem[\protect\citeauthoryear{Hellings}{1986}]{Hellings1986} Hellings R.~W., 1986, AJ, 91, 650. doi:10.1086/114048

\bibitem[\protect\citeauthoryear{Kellermann et al.}{1998}]{Kellermann1998} Kellermann K.~I., Vermeulen R.~C., Zensus J.~A., Cohen M.~H., 1998, AJ, 115, 1295. doi:10.1086/300308

\bibitem[\protect\citeauthoryear{Klioner}{1991}]{Klioner1991} Klioner S.~A., 1991, gvmg.conf, 188

\bibitem[\protect\citeauthoryear{Kopeikin}{1990}]{Kopeikin1990} Kopeikin S.~M., 1990, SvA, 34, 5

\bibitem[\protect\citeauthoryear{Ma et al.}{1998}]{Ma1998} Ma C., Arias E.~F., Eubanks T.~M., Fey A.~L., Gontier A.-M., Jacobs C.~S., Sovers O.~J., et al., 1998, AJ, 116, 516. doi:10.1086/300408

\bibitem[\protect\citeauthoryear{Mansouri \& Sexl}{1977}]{Mansouri1977} Mansouri R., Sexl R.~U., 1977, GReGr, 8, 497. doi:10.1007/BF00762634

\bibitem[\protect\citeauthoryear{Petit \& Luzum}{2010}]{iers10} Petit G., Luzum B. (eds.), 2010, IERS Technical Notes, 36, 1

\bibitem[\protect\citeauthoryear{Schuh \& Behrend}{2012}]{Schuh12} Schuh H., Behrend D., 2012, JGeo, 61, 68. doi:10.1016/j.jog.2012.07.007

\bibitem[\protect\citeauthoryear{Soffel et al.}{1991}]{Soffel1991} Soffel M.~H., Wu X., Xu C., Mueller J., 1991, AJ, 101, 2306. doi:10.1086/115851

\bibitem[\protect\citeauthoryear{Soffel, Kopeikin, \& Han}{2017}]{2017JGeod..91..783S} Soffel M., Kopeikin S., Han W.-B., 2017, JGeod, 91, 783. doi:10.1007/s00190-016-0956-z

\bibitem[\protect\citeauthoryear{Titov, Tesmer, \& Boehm}{2004}]{Titov2004} Titov O., Tesmer V., Boehm J., 2004, ivsg.conf, 267

\bibitem[\protect\citeauthoryear{Titov \& Girdiuk}{2015}]{2015A&A...574A.128T} Titov O., Girdiuk A., 2015, Astron.Astroph., 574, A128. doi:10.1051/0004-6361/201424690

\bibitem[\protect\citeauthoryear{Titov \& Kr{\'a}sn{\'a}}{2019}]{Titov2019} Titov O., Kr{\'a}sn{\'a} H., 2019, In: {FreyMueller} J, {S{\'a}nchez} L. (eds),
  International Symposium on Advancing Geodesy in a Changing World, International Association of Geodesy Symposia, vol 149. Springer, Cham, 19, arXiv:1808.06769

\bibitem[\protect\citeauthoryear{Titov \& Kr{\'a}sn{\'a}}{2018}]{2018A&A...610A..36T} Titov O., Kr{\'a}sn{\'a} H., 2018, Astron.Astroph., 610, A36. doi:10.1051/0004-6361/201731901
                                                                                                                             
\bibitem[\protect\citeauthoryear{Whitney}{2000}]{2000ivsg.conf..187W} Whitney, A.~R., 2000, International VLBI Service for Geodesy and Astrometry 2000 General Meeting Proceedings, 187.

\bibitem[\protect\citeauthoryear{Will}{1971}]{Will71} Will C.~M., 1971, ApJ, 163, 611. doi:10.1086/150804

\bibitem[\protect\citeauthoryear{Will}{1992}]{Will1992} Will C.~M., 1992, PhRvD, 45, 403. doi:10.1103/PhysRevD.45.403


\end{thebibliography}

\clearpage

\begin{appendix}

\section{DEVELOPMENT OF THE RELATIVISTIC GROUP DELAY MODELS FOR THE EPOCH OF GEOCENTER AND FOR THE EPOCH OF THE REFERENCE STATION}

\subsection{THE CONVENTIONAL GEOCENTRIC DELAY MODEL}
The equation for the relativistic group delay model has been developed in the 1980s-90s (e.g. \cite{Hellings1986}, \cite{Kopeikin1990}, \cite{Klioner1991}, \cite{Soffel1991} to approximate the observed VLBI data at the 1-ps level of accuracy. The conventional group delay model was finally adopted \cite{iers10}

\begin{equation}\label{groupdelay_gcrs}
\tau_{g}=\frac{-\frac{(\boldsymbol{b}\cdot\boldsymbol{s})}{\textrm{c}}\Big(1-\frac{2GM}{c^2R}
-\frac{|\boldsymbol{V}|^{2}}{2\textrm{c}^{2}}
-\frac{(\boldsymbol{V}\cdot\boldsymbol{w_2})}{\textrm{c}^{2}}\Big)
-\frac{(\boldsymbol{b}\cdot\boldsymbol{V})}{\textrm{c}^{2}}\Big(1+\frac{(\boldsymbol{s}\cdot\boldsymbol{V})}{2\textrm{c}}\Big)}
    {1+\frac{(\boldsymbol{s}\cdot(\boldsymbol{V}+{w_2}))}{\textrm{c}}}
\end{equation}

where $\boldsymbol{b}$ is the vector of baseline $\boldsymbol{b} = \boldsymbol{r}_{2} - \boldsymbol{r}_{1}$,  $\boldsymbol{s}$ is the barycentric unit vector of the radio source, $\boldsymbol{V}$ is the barycentric velocity of the geocenter, $\boldsymbol{w_2}$ is the geocentric velocity of station 2, $c$ is the speed of light, $G$ is the gravitational constant, $M$ is the mass of the Sun, $R$ is the geocentric distance to the Sun, and ($\cdot$) is the dot-product operator of two vectors. The reference epoch is the UTC epoch of the wavefront passage at the reference station. In accordance with the assumption, station 1 is treated as the reference station, the geocentric velocity of station 2 is presented in (1) explicitly. A modern revision (e.g. \cite{2017JGeod..91..783S}) is to add some smaller terms (less than 1 ps), but the analytical model (1) is still valid for the analysis of VLBI data.

\subsection{LORENTZ TRANSFORMATION}

The radio signal is received by two radio telescopes on the surface of the rotating Earth, and their coordinates are presented in the Geocentric Celestial Reference System (GCRS) comoving with the Earth. Positions of reference radio sources emitting the signals are in the Barycentric Celestial Reference System (BCRS). So, a detailed transformation of the coordinates from BCRS to GCRS is traditionally based on the metric tensor of the Solar System at the first and second post-Newtonian level (e.g. \cite{Hellings1986},  \cite{Kopeikin1990}, \cite{Klioner1991}, \cite{2017JGeod..91..783S}). However, many lower order effects are not observable, therefore, a simplified approach could be developed for the relativistic model delay.

The conventional Lorenz transformation is given by

\begin{equation}\label{xt}
\begin{aligned}
\boldsymbol{x'} = & \boldsymbol{x}+(\gamma-1)\frac{(\boldsymbol{V}\cdot\boldsymbol{x})\boldsymbol{V}}{|\boldsymbol{V}|^2} - \gamma\boldsymbol{V}t \\
t' = & \gamma\Bigg(t - \frac{(\boldsymbol{V}\cdot\boldsymbol{x})}{\textrm{c}^2}\Bigg) .
\end{aligned}
\end{equation}

where $\gamma = \bigg(\sqrt{1-\frac{|\boldsymbol{V}|^2}{\textrm{c}^2}}\bigg)^{-1}$ is the Lorentz "gamma-factor" (\cite{Mansouri1977}, \cite{Will1992}). It should be noted that this factor here is not the parameter $\gamma$ of the Post-Newtonian formalism (PPN) used in general relativity \cite{Will71}.

Transformation (\ref{xt}) links the geocentric reference system $S'(x',t')$ that is moving with velocity $\boldsymbol{V}$ around the Solar System Barycentre (SSB) and the barycentric reference system $S(x,t)$ located at the SSB. 
It could be shown (\cite{Titov2019}) that the time delay derived from (\ref{xt}) may be presented in the form

\begin{equation}\label{groupdelay_gcrs1}
\tau_{g_0}=\frac{-\frac{(\boldsymbol{b}\cdot\boldsymbol{s})}{\textrm{c}}\Big(1-\frac{|\boldsymbol{V}|^{2}}{2\textrm{c}^{2}}\Big) -\frac{(\boldsymbol{b}\cdot\boldsymbol{V})}{\textrm{c}^{2}}\Big(1+\frac{(\boldsymbol{s}\cdot\boldsymbol{V})}{2\textrm{c}}\Big)}
    {1+\frac{(\boldsymbol{s}\cdot\boldsymbol{V})}{\textrm{c}}}
\end{equation}

Whether an astronomical instrument with a reference clock were placed in the Earth's geocenter and the Solar gravitation were ignored, the equation (\ref{groupdelay_gcrs1}) would be applied to reduction of the geodetic VLBI data. However, further complications will be discussed in two next subsections.

\subsubsection{SPACE AND TIME TRANSFORMATION INCLUDING GRAVITATIONAL POTENTIAL}

The relativistic model (\ref{groupdelay_gcrs1}) does not include the term proportional to the Solar gravitational potential $\frac{2U}{c^2}$, where $U = \frac{GM}{R}$, and few terms with the geocentric velocity $\boldsymbol{w_2}$ presented in (\ref{groupdelay_gcrs}). \cite{Hellings1986} showed that the former term appears due to the Solar gravitational field (in the Schwarzschild metric) at the Earth geocentre. Therefore, \cite{Hellings1986} has developed new equations for the relationships between intervals of physical distance and time, measured in a moving
reference geocentric frame, and the intervals, given in the barycentric coordinate system including the gravitational field of the Sun is given by

\begin{equation}\label{xt1}
\begin{aligned}
\boldsymbol{x'} = & (1+\frac{2U}{c^2})\boldsymbol{x}-(1+\frac{2U}{c^2})(\gamma-1)\frac{(\boldsymbol{V}\cdot\boldsymbol{x})\boldsymbol{V}}{|\boldsymbol{V}|^2}- \\
& - (1-\frac{2U}{c^2})\gamma\boldsymbol{V}t \\
t' = & \gamma\Bigg((1-\frac{2U}{c^2})t - (1+\frac{2U}{c^2})\frac{(\boldsymbol{V}\cdot\boldsymbol{x})}{\textrm{c}^2}\Bigg) .
\end{aligned}
\end{equation}

Transformation (\ref{xt1}) reduces to the Lorenz transformation (\ref{xt}) if the Solar potential $U=0$.

The corresponding equation for the relativistic group delay includes the Solar gravitational potential at the geocenter of the Earth. 

\begin{equation}\label{groupdelay_gcrs2}
\tau_{g_U}=\frac{-\frac{(\boldsymbol{b}\cdot\boldsymbol{s})}{\textrm{c}}\Big(1-\frac{2U}{c^2}-\frac{|\boldsymbol{V}|^{2}}{2\textrm{c}^{2}}\Big) -\frac{(\boldsymbol{b}\cdot\boldsymbol{V})}{\textrm{c}^{2}}\Big(1+\frac{(\boldsymbol{s}\cdot\boldsymbol{V})}{2\textrm{c}}\Big)}
    {1+\frac{(\boldsymbol{s}\cdot\boldsymbol{V})}{\textrm{c}}}
\end{equation}

\cite{2015A&A...574A.128T} showed that the term proportional to $\frac{2U}{c^2}$ in (\ref{groupdelay_gcrs2}) could be unified with the general relativity effect of the gravitational delay. Therefore, we will not include it into further analysis; however, we discuss it here as it is a part of the conventional geometric part of the relativistic delay model (\cite{iers10}).

\subsubsection{LORENZ TRANSFORMATION REFERRING TO THE EPOCH OF FIRST STATION}

 Physical clocks (hydrogen masers) used for VLBI observations are located at the Earth surface rather than at the geocenter. As two clocks separated by a long baseline are involved for a routine observational experiment, one of them should be selected as "reference" clock. This choice is completely arbitrary, though, once it is made, the geocentric velocity of the second ("no reference") clock appears explicitly in the analytical equations. The standard approach is to consider a difference between barycentric coordinates of two radio telescopes, $\boldsymbol{r_1}(t_1)$ and $\boldsymbol{r_2}(t_2)$, measured at the two epochs ${t_1}$ and ${t_2}$, to expand the vector $\boldsymbol{r_2}(t_2)$ as follows

\begin{equation}\label{xt2}
\begin{aligned}
\boldsymbol{r_2}(t_2) = \boldsymbol{r_2}(t_1) + \boldsymbol{w_2(t_1)}(t_2 - t_1),
\end{aligned}
\end{equation}

where $\boldsymbol{w_2}=\boldsymbol{w_2(t_1)}$ is the geocentric velocity of the second station at epoch $t_1$. Denoting $\boldsymbol{B(t_1)}$ a difference between two barycentric vectors at the same epoch $\boldsymbol{B} = \boldsymbol{B(t_1)} = \boldsymbol{r_2(t_1)} - \boldsymbol{r_1(t_1)}$ one could get for the time difference $ (t_2 - t_1)$ 

\begin{equation}\label{ct}
\begin{aligned}
c(t_2 - t_1) = - (\boldsymbol{B}\cdot\boldsymbol{s})-(\boldsymbol{w_2\cdot\boldsymbol{s}})(t_2 - t_1)
\end{aligned}
\end{equation}

It should be noted here that $\boldsymbol{B}$ is a formal three-component vector rather than a meaningful physical value, though it links to the physical distance between two terrestrial positions of radio telescopes on the Earth at  ${t_1}$.


Eq (\ref{ct}) could be obtained by alternative way. Let's introduce of a new geocentric reference frame $S'' = S''(x'',t'')$  with the reference epoch referred to station 1 in a such way that two geocentric reference frames $S''$ and $S'$ are linked by new transformation

\begin{equation}\label{w_2}
\begin{aligned}
\boldsymbol{x"} = & \boldsymbol{x'} \\
t" = & t' - \frac{(\boldsymbol{w_2}\cdot\boldsymbol{x'})}{\textrm{c}^{2}}
\end{aligned}
\end{equation}

Transformation (\ref{w_2}) could be easily combined with the Lorentz transformation (\ref{xt})

\begin{equation}\label{xt3}
\begin{aligned}
\boldsymbol{x"} = & \boldsymbol{x}+(\gamma-1)\frac{(\boldsymbol{V}\cdot\boldsymbol{x})\boldsymbol{V}}{|\boldsymbol{V}|^2} - \gamma\boldsymbol{V}t \\
t" = & \gamma\Bigg(t - \frac{(\boldsymbol{V}\cdot\boldsymbol{x})}{\textrm{c}^2}\Bigg)
- \frac{(\boldsymbol{w_2}\cdot\boldsymbol{x})}{\textrm{c}^2} - \\  
& -(\gamma-1)\frac{(\boldsymbol{V}\cdot\boldsymbol{x})(\boldsymbol{V}\cdot\boldsymbol{w_2})}{{c^2}\cdot|\boldsymbol{V}|^2}+
\gamma\frac{(\boldsymbol{V}\cdot\boldsymbol{w_2})t}{\boldsymbol{c}^2}.
\end{aligned}
\end{equation}

 It is obvious that the transformations (\ref{w_2}) and (\ref{xt3}) are pertinent only for an individual pair of two radio telescopes equipped with their own high precision clocks, one of which is a reference clock and the second clock is moving with instantaneous velocity $\boldsymbol{w_2}$ . The transformation (\ref{xt3}) is fully consistent with the special relativity postulates and reflects the situation when the position of the reference clock is not at the reference frame origin (geocentre). For a classical astronomic instrument the reference frame origin and position of the reference clock are referred to the same topocentric position of the instrument on the Earth surface. In this scenario, the geocentric velocity of the instrument is simply added to the barycentric velocity in the formulae of the Lorenz transformation, i.e. the velocity $\boldsymbol{V}$ is replaced by the sum $\boldsymbol{V} + \boldsymbol{w_2}$ in (\ref{xt}) followed by a substantial change in (\ref{groupdelay_gcrs1}). 
From the observational point of view this results in the appearance of the classical diurnal aberration effect. In geodetic VLBI, there is no the diurnal aberration effect at all. Instead of that, as it will be shown later, the geocentric velocity $\boldsymbol{w_2}$ contributes to the diurnal variation of the scale factor with magnitude up to 20 ns (or 6 meters in the linear scale) for a standard baseline of 6000 km in length and a geocentric velocity of 300 m/s. 
 
For calculating the group delay from (\ref{xt3}), one needs to develop the corresponding velocity transformation. As both reference frames $S''$ and $S'$ are geocentric, the time component is only changed due to transition from (\ref{xt}) to (\ref{xt3}). Traditionally, authors proceed to the equation of the relativistic time delay (\ref{groupdelay_gcrs1}) consistent with the XF-type correlator directly (e.g. Hellings 1986, Kopeikin 1990, Soffel et al 2017). Therefore, these two transformations (\ref{xt}) and (\ref{w_2})  merge together and the difference between the delays (\ref{groupdelay_gcrs}) and (\ref{groupdelay_gcrs1}) is lost. However, for the FX-type correlators, this procedure must be separated into two steps to provide a proper relativistic conversion between observables produced by the XF and FX correlators.

To elaborate equation (\ref{groupdelay_gcrs}) (without the $\frac{2U}{c^2}$ term) from transformation (\ref{xt3}), let's consider the velocity transformation ${\boldsymbol{v_x"}}=\frac{\boldsymbol{dx"}}{dt"}$

\begin{equation}\label{vx1}
\begin{aligned}
{v_x"}=\frac{\boldsymbol{dx}+(\gamma-1)\frac{(\boldsymbol{V}\cdot\boldsymbol{dx})\boldsymbol{V}}{|\boldsymbol{V}|^2}-\gamma\boldsymbol{V}dt}
{\gamma\Bigg((1+\frac{(\boldsymbol{V}\cdot\boldsymbol{w_2})}{\boldsymbol{c}^2})dt- \frac{(\boldsymbol{V}\cdot\boldsymbol{dx})}{\textrm{c}^2}\Bigg)
- \frac{(\boldsymbol{w_2}\cdot\boldsymbol{dx})}{\textrm{c}^2} - \frac{(\boldsymbol{V}\cdot\boldsymbol{dx})(\boldsymbol{V}\cdot\boldsymbol{w_2})}{2{c}^4}}
\end{aligned}
\end{equation}

or, denoting ${\boldsymbol{v_x}}=\frac{\boldsymbol{dx}}{dt}$ within the 1 ps level of accuracy

\begin{equation}\label{vx2}
\begin{aligned}
{v_x"}=\frac{\boldsymbol{v_x}+\frac{(\boldsymbol{V}\cdot\boldsymbol{v_x})\boldsymbol{V}}{2c^2}-\boldsymbol{V}}{\Bigg(1+\frac{(\boldsymbol{V}\cdot\boldsymbol{w_2})}{\boldsymbol{c}^2}-\frac{(\boldsymbol{V}\cdot\boldsymbol{v_x})}{\textrm{c}^2}\Bigg)
\Bigg(1+\frac{|\boldsymbol{V}|^2}{2c^2}\Bigg)
- \frac{(\boldsymbol{w_2}\cdot{\boldsymbol{v_x}})}{\textrm{c}^2}}
\end{aligned}
\end{equation}

Now apply for a standard transition to the radio source vector $c\boldsymbol{s} = -\boldsymbol{v_x}$

\begin{equation}\label{s""1}
\begin{aligned}
{s"}=\frac{{s}+\frac{(\boldsymbol{V}\cdot\boldsymbol{s})\boldsymbol{V}}{2c^2}+\frac{\boldsymbol{V}}{c}}{\Bigg(1+\frac{(\boldsymbol{V}\cdot\boldsymbol{s})}{c}+\frac{(\boldsymbol{V}\cdot\boldsymbol{w_2})}{c^2}\Bigg)\Bigg(1+\frac{|\boldsymbol{V}|^2}{2c^2}\Bigg)+\frac{(\boldsymbol{w_2}\cdot{s})}{\textrm{c}}}
\end{aligned}
\end{equation}

and, after reduction of negligible terms,

\begin{equation}\label{s""2}
\begin{aligned}
{s"}=\frac{{s}+\frac{(\boldsymbol{V}\cdot\boldsymbol{s})\boldsymbol{V}}{2c^2}+\frac{\boldsymbol{V}}{c}}{1+\frac{(\boldsymbol{V}\cdot\boldsymbol{s})}{c}+\frac{(\boldsymbol{V}\cdot\boldsymbol{w_2})}{c^2}+\frac{|\boldsymbol{V}|^2}{2c^2}+\frac{(\boldsymbol{w_2}\cdot{s})}{\textrm{c}}}
\end{aligned}
\end{equation}

This equation could be converted to the form consistent with the conventional group delay model at 1-ps level after inclusion of the Solar gravitation term (\ref{groupdelay_gcrs2})

\begin{equation}\label{s""3}
\begin{aligned}
{s"}=\frac{{s}\Bigg(1-\frac{2U}{c^2}-\frac{|\boldsymbol{V}|^2}{2c^2}-\frac{(\boldsymbol{V}\cdot\boldsymbol{w_2})}{c^2}\Bigg)+\frac{\boldsymbol{V}}{c}\Bigg(1+\frac{(\boldsymbol{V}\cdot\boldsymbol{s})}{2c}\Bigg)}{1+\frac{((\boldsymbol{V}+\boldsymbol{w_2})\cdot\boldsymbol{s})}{c}}
\end{aligned}
\end{equation}

\hspace{3mm}

Development of the time delay from (\ref{s""3}) as $ \tau = -\frac{(\boldsymbol{b}\cdot\boldsymbol{s"})}{c}$ provides the conventional group delay model (\ref{groupdelay_gcrs}).  Now it is obvious that this model is based on the modification of the Lorentz transformation (\ref{xt3}) in which the transformation of time is presented in a non-standard way because our reference clocks are physically located at the Earth surface rather than at the geocenter.

Eq (\ref{groupdelay_gcrs1}) misses the terms including the velocity of the second radio telescope in (\ref{groupdelay_gcrs}).  At the 1 ps level of accuracy this difference $\delta\tau = \tau_g - \tau_{g_0}$  comprises five terms

\begin{equation}\label{delta_tau}
\begin{aligned}
\delta\tau = \frac{2(\boldsymbol{b}\cdot\boldsymbol{s}){U}}{c^3} + \frac{(\boldsymbol{b}\cdot\boldsymbol{s}){(\boldsymbol{w_2}\cdot\boldsymbol{s})}}{c^2}+\\
+ \frac{(\boldsymbol{b}\cdot\boldsymbol{s}){(\boldsymbol{V}\cdot\boldsymbol{w_2})}}{c^3}
+\frac{(\boldsymbol{b}\cdot\boldsymbol{V}){(\boldsymbol{w_2}\cdot\boldsymbol{s})}}{c^3} - \\
-\frac{2(\boldsymbol{b}\cdot\boldsymbol{s}){(\boldsymbol{V}\cdot\boldsymbol{s})}{(\boldsymbol{w_2}\cdot\boldsymbol{s})}}{c^3} 
\end{aligned}
\end{equation}

\end{appendix}

\clearpage

\begin{table}[ht]
\centering 
\caption{Estimates of the parameter $\epsilon$ (units $10^{-3}$) for six VGOS stations in 2019)}
\resizebox{\textwidth}{!}{\begin{tabular}{lrrrrrr} \hline
MJD	   &  GGAO12M	       	& KOKEE12M &	ONSA13NE &	ONSA13SW &	RAEGYEB &	WESTFORD  \\
\hline
58534.2499 & $0.015 \pm 0.049 $   & $ 0.055 \pm 0.045 $   & $ 0.164 \pm 0.053 $   &   --------------                & $ 0.127 \pm 0.051 $  &   $-0.135 \pm 0.047$ \\    
58547.2499 & $-0.186 \pm 0.049$   & $ 0.265 \pm 0.043  $   & $ 0.258 \pm 0.053 $   &   --------------                & $ 0.173 \pm 0.053 $  &   $-0.337 \pm 0.049$ \\   
58561.2499 & $-0.425 \pm 0.081$   & $ 0.051 \pm 0.060  $   & $ 0.243 \pm 0.077 $   &   --------------                & $ 0.106 \pm 0.068 $  &   $-0.150 \pm 0.060$ \\   
58575.2500 & $0.016 \pm 0.054 $   & $ 0.078 \pm 0.047 $   & $ 0.117 \pm 0.057 $   &   --------------                & $ 0.312 \pm 0.054 $  &   $-0.042 \pm 0.064$ \\   
58589.2497 & $-0.252 \pm 0.061 $   & $ 0.262 \pm 0.050  $   & $ 0.233 \pm 0.055 $   &   --------------                &  --------------             &   $-0.278 \pm 0.057$ \\   
58603.2497 & $-0.330 \pm 0.062$   & $ 0.290 \pm 0.051  $   & $ 0.221 \pm 0.058 $   &   --------------                &  --------------             &   $-0.333 \pm 0.061$ \\   
58617.2497 & $-0.109 \pm 0.058$   & $ 0.117 \pm 0.051 $   & $ 0.038 \pm 0.065 $   &   --------------                &  --------------             &   $-0.097 \pm 0.062$ \\   
58632.2494 & $-0.563 \pm 0.122 $   &  --------------               & $ -0.202 \pm 0.074 $   &   --------------                &  --------------             &    --------------           \\   
58659.2497 & $0.049 \pm 0.076 $   & $ 0.058 \pm 0.061  $   & $ 0.166 \pm 0.067 $   &   --------------                &  --------------             &   $-0.178 \pm 0.069$ \\   
58673.2499 & $-0.085 \pm 0.180 $   &  --------------               & $ -0.068 \pm 0.107 $   &   --------------                &  --------------             &    --------------           \\   
58687.2496 & $0.015 \pm 0.080 $   & $ -0.068 \pm 0.091   $ & $ 0.008 \pm 0.077 $   &  $  0.001 \pm 0.076 $   &  --------------             &   $ 0.024 \pm 0.066$ \\   
58701.2499 & $-0.068 \pm 0.068 $   & $ 0.069 \pm 0.059   $ & $ 0.255 \pm 0.074 $   &  $  0.223 \pm 0.074 $   & $ 0.155 \pm 0.072  $ &   $-0.237 \pm 0.060 $ \\   
58715.2498 & $-0.282 \pm 0.050$   &  		 			   & $ 0.066 \pm 0.055 $   &  $  0.068 \pm 0.054 $   & $ 0.145 \pm 0.059  $ &   $0.025 \pm 0.049 $ \\   
58732.2499 & $-0.160 \pm 0.047$   & $ 0.099 \pm 0.041    $ & $ -0.002 \pm 0.048 $   &  $  -0.037 \pm 0.048 $   & $ 0.056 \pm 0.050 $ &   $-0.025 \pm 0.039$ \\   
58743.2499 & $-0.206 \pm 0.045$   & $ 0.128 \pm 0.038    $ & $ 0.204 \pm 0.046 $   &  $  0.153 \pm 0.045  $   & $ 0.311 \pm 0.046  $ &   $-0.124 \pm 0.036$ \\   
58757.2497 & $-0.171 \pm 0.053$   & $ 0.147 \pm 0.042   $ & $ 0.173 \pm 0.048$   &  $  0.226 \pm 0.050  $   &  --------------             &   $-0.145 \pm 0.044 $ \\   
58774.2493 & $-0.885 \pm 0.071$   & $ 0.892 \pm 0.079    $ & $ -0.220 \pm 0.072$   &  $  -0.128 \pm 0.073  $   &  --------------             &    --------------           \\   
58785.2496 & $-0.214 \pm 0.051 $   & $ 0.194 \pm 0.031   $ & $ 0.153 \pm 0.038 $   &  $  0.143 \pm 0.038  $   &  --------------             &    --------------           \\   
58802.2498 & $0.106 \pm 0.062 $   & $ -0.034 \pm 0.035   $ & $ -0.007 \pm 0.047$   &  $  -0.004 \pm 0.048 $   &  --------------             &    --------------           \\   
58813.2500 & $-0.135 \pm 0.046$   & $ 0.150 \pm 0.036    $ & $ 0.206 \pm 0.048 $   &  $  0.195 \pm 0.048  $   &  --------------             &   $-0.189 \pm 0.044$ \\   
58827.2497 & $-0.211 \pm 0.056$   & --------------                & $ 0.182 \pm 0.042 $   &  $  0.183 \pm 0.042  $   &  --------------             &   $-0.135 \pm 0.041$ \\   
58844.2499 & --------------              & $ 0.472 \pm 0.073    $ & $ -0.018 \pm 0.074 $   &  $  -0.148 \pm 0.074 $   &  --------------             &   $-0.223 \pm 0.049$ \\   
58858.2498 & $-0.072 \pm 0.066$   & --------------                & $ -0.123 \pm 0.051 $   &  $  -0.058 \pm 0.052  $   &  --------------             &   $0.048 \pm 0.047$ \\     
\hline                                                          
\end{tabular}}
\end{table}

\end{document}